\documentclass[manuscript]{acmart}

\PassOptionsToPackage{hyphens}{url}
\usepackage{url}
\usepackage{soul}
\usepackage{xcolor}
\usepackage{comment}
\usepackage{enumitem}
\usepackage{subcaption}
\usepackage{siunitx}

\copyrightyear{2022}
\acmYear{2022}
\setcopyright{rightsretained}
\acmConference[HT '22]{Proceedings of the 33rd ACM Conference on Hypertext and Social Media}{June 28-July 1, 2022}{Barcelona, Spain}
\acmBooktitle{Proceedings of the 33rd ACM Conference on Hypertext and Social Media (HT '22), June 28-July 1, 2022, Barcelona, Spain}
\acmDOI{10.1145/3511095.3536369}
\acmISBN{978-1-4503-9233-4/22/06}

\begin{document}

\title{Personalized Interventions for Online Moderation}

\author{Stefano Cresci}
\email{stefano.cresci@iit.cnr.it}
\orcid{0003-0170-2445}
\affiliation{%
    \institution{Institute for Informatics and Telematics, National Research Council (IIT-CNR)}
    \streetaddress{via G. Moruzzi 1}
    \city{Pisa}
    \country{Italy}
    \postcode{56124}
}

\author{Amaury Trujillo}
\email{amaury.trujillo@iit.cnr.it}
\affiliation{%
    \institution{Institute for Informatics and Telematics, National Research Council (IIT-CNR)}
    \streetaddress{via G. Moruzzi 1}
    \city{Pisa}
    \country{Italy}
    \postcode{56124}
}

\author{Tiziano Fagni}
\email{tiziano.fagni@iit.cnr.it}
\affiliation{%
    \institution{Institute for Informatics and Telematics, National Research Council (IIT-CNR)}
    \streetaddress{via G. Moruzzi 1}
    \city{Pisa}
    \country{Italy}
    \postcode{56124}
}

\renewcommand{\shortauthors}{Cresci et al.}

\begin{abstract}
Current online moderation follows a one-size-fits-all approach, where each intervention is applied in the same way to all users. This na\"{i}ve approach is challenged by established socio-behavioral theories and by recent empirical results that showed the limited effectiveness of such interventions. We propose a paradigm-shift in online moderation by moving towards a personalized and user-centered approach. Our multidisciplinary vision combines state-of-the-art theories and practices in diverse fields such as computer science, sociology and psychology, to design \textit{personalized moderation interventions} (PMIs). In outlining the path leading to the next-generation of moderation interventions, we also discuss the most prominent challenges introduced by such a disruptive change.
\end{abstract}


\maketitle




\section{Introduction}
\label{sec:introduction}

Nowadays, social media play a pivotal role in shaping public opinion. Online users constantly read, create, and share an ever-growing amount of content, with social media now overtaking the more traditional media, especially among the younger generations. On the one hand, this paradigm-shift created new opportunities for civic engagement and for democratizing access to information~\cite{tucker2017liberation}. On the other hand, this freedom also gave rise to multiple online harms, such as misinformation, polarization, toxic and hateful speech~\cite{dipietro2021new}. The consequences of such harms are not limited to online platforms, but also affect the offline world, as demonstrated by recent political riots~\cite{eip2021fuse} and by the decreased confidence in vaccines~\cite{loomba2021measuring}.

For this reason, since the 2016 Donald Trump presidential win and the UK Brexit referendum, platforms have been facing a \textit{tremendous public and governmental pressure} to take action against online harms. Recent dramatic events such as the COVID-19 infodemic and the Russian-Ukrainian conflict increased such pressure even more. 
Platforms responded to the growing pressure by hastily deploying a number of \textit{moderation interventions} ---actions taken to enforce content policies and rules. For example, Twitter, Facebook and Instagram attached warning labels to disputed posts~\cite{zannettou2021won} and banned users, groups, and pages that misbehaved~\cite{jhaver2021evaluating}. Pinterest blocked search results for anti-vaccination queries and Reddit quarantined and banned toxic communities~\cite{saleem2018aftermath,horta2021platform,trujillo2022make}. However, despite appearing as reasonable solutions and serving as public evidence of the platforms’ willingness to tackle the issues they contributed to create, these interventions were designed and applied light-mindedly. Recent studies measured limited or no effects at all~\cite{chandrasekharan2020quarantined,horta2021platform,trujillo2022make}, and showed that some interventions even exacerbated the very issues they aimed to solve~\cite{zollo2017debunking,bail2018exposure,dias2020emphasizing,pennycook2020implied}. Overall, the design of moderation interventions received limited scholarly attention and progress was sought by trial-and-error rather than by a rigorous scientific approach. Pennycook \& Rand concluded an op-ed\footnote{\url{https://www.nytimes.com/2020/03/24/opinion/fake-news-social-media.html}} on The New York Times remarking that moderation interventions should «not just rely on common sense or intuition» but that should instead be «empirically grounded». Science-based online moderation is still in its infancy.

\section{The case for personalization in online moderation}
\label{sec:personalization}

Until now, online moderation always followed a \textit{one-size-fits-all} approach, where each intervention was applied in the same way for all users. As a recent example, when Twitter sent warning messages to dissuade users from posting toxic tweets, all users received the exact same message~\cite{katsaros2021reconsidering}. However, results on user reactions to moderation interventions showed that different users react in different ways to the same intervention, according to their individual characteristics~\cite{saleem2018aftermath,trujillo2022make}. Theories from the social, psychological, and behavioral sciences support these empirical results~\cite{drkazkiewicz2022study} and posit that the efficacy of interventions depends on individual and contextual characteristics~\cite{williams2017individual}. In line with this literature, but in contrast to the platforms' objectives, the interventions recently applied by Twitter and Reddit against toxic users caused a subset of users to become even more toxic and radicalized~\cite{katsaros2021reconsidering,horta2021platform,trujillo2022make}. 


\begin{figure*}[t]
    \centering
    \includegraphics[width=0.9\textwidth]{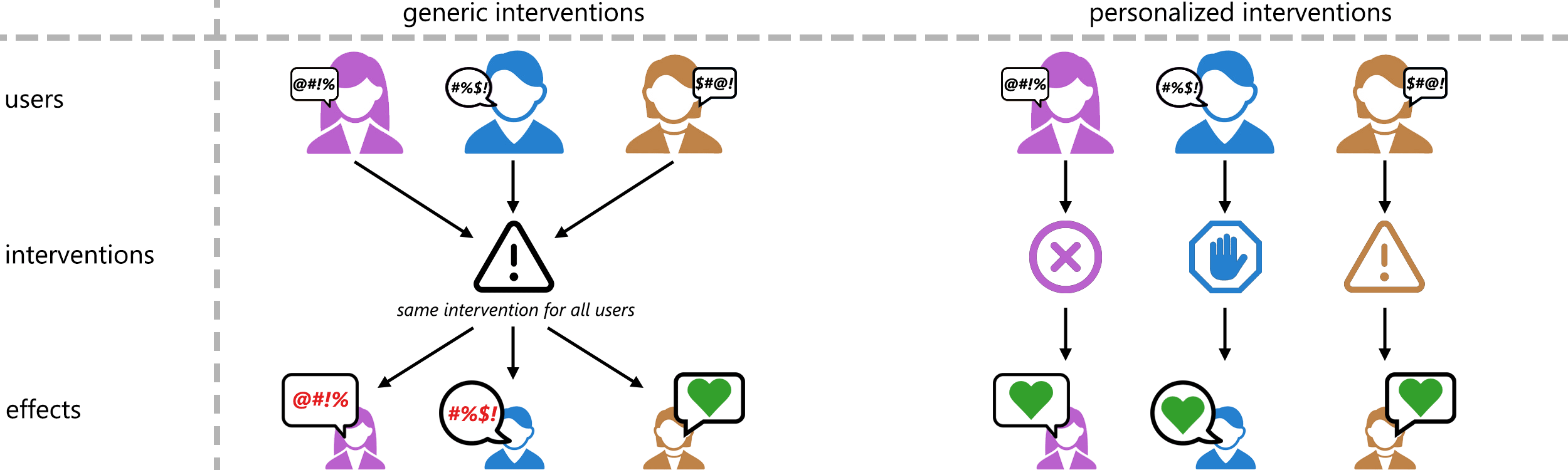}
    \caption{Contrarily to the common practice of generic interventions, personalized moderation interventions are tailored to the individual characteristics of the moderated users. This personalization allows maximizing the desired effects of the moderation.}
    \label{fig:generic-vs-personalized}
\end{figure*}

Overall, the existing literature across multiple disciplines opposes the current one-size-fits-all approach to online moderation and instead suggests that \textit{interventions should be tailored to the individual characteristics of the users}. In other words, the current na\"{i}ve approach to online moderation neglects individual differences and the advantages of \textit{personalization}. Indeed, personalization has already proved valuable in several online domains, such as advertising, music and video streaming services, and the improvement of health-related behavior via apps~\cite{jacobson2016music,wang2020just}. By taking inspiration from medicine ---a field with which online moderation shares many commonalities--- we observe that current generic interventions are deployed as a sort of \emph{universal cure} to treat the ailments of all online users, instead of following the virtuous examples of \emph{personalized medicine} where patients affected by a disease receive personalized treatment based on their condition, individual characteristics, environment, and behavior.


\section{Towards personalized moderation interventions}
\label{sec:personalized-interventions}

\begin{figure*}[t]
    \centering
    \includegraphics[width=0.9\textwidth]{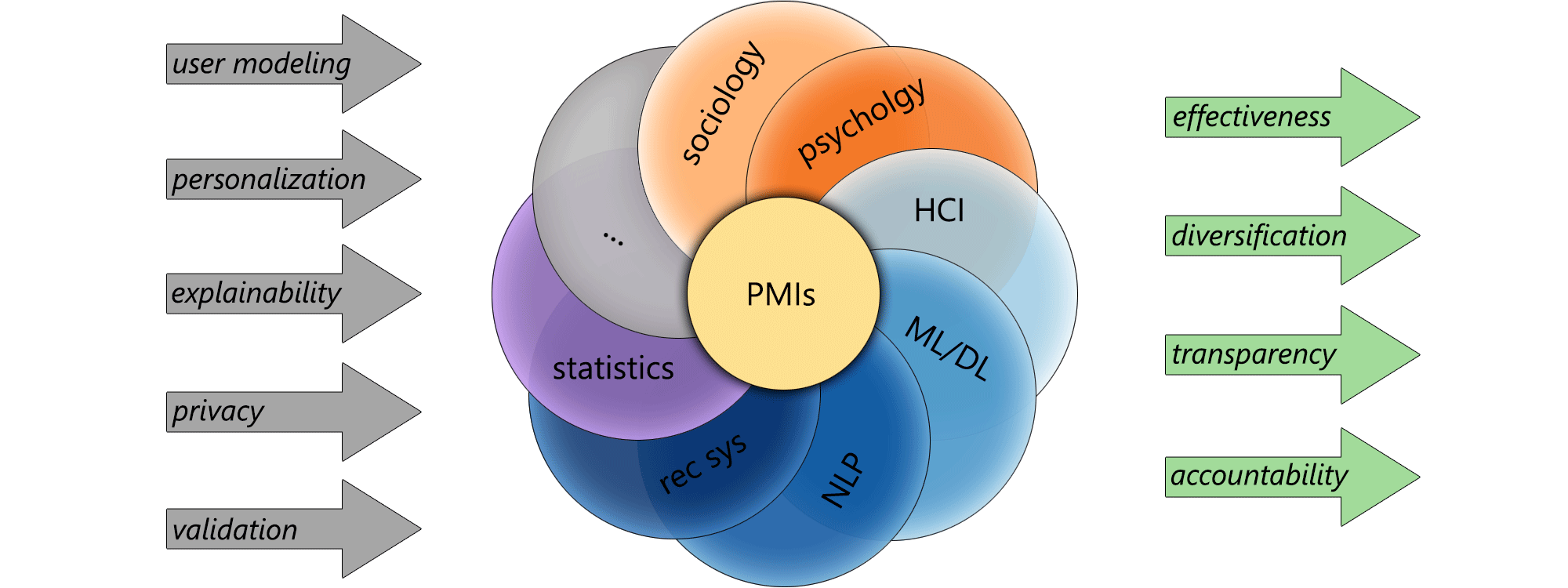}
    \caption{Developing personalized moderation interventions (PMIs) will involve conjoint efforts by multiple scientific communities, including sociology, psychology and computer science.}
    \label{fig:sci-areas}
\end{figure*}

Times are ripe for a disruptive change in the approach to online moderation. Motivated by the aforementioned empirical results showing the inadequacy of current generic interventions, the socio-behavioral theories that support the application of personalized interventions, and the proven benefits of personalization in many domains, we envision a paradigm-shift towards \emph{personalized moderation interventions} (PMIs). As depicted in Figure~\ref{fig:generic-vs-personalized}, PMIs are tailored to the individual characteristics of the users to which they are applied, which allows maximizing the desired effects of the moderation, while minimizing the possible undesired side effects. To fulfill the vision of PMIs, existing knowledge and methods from several disciplines and scientific communities must be combined in order to produce new knowledge and the methods needed to develop PMIs. As illustrated in Figure~\ref{fig:sci-areas}, among the scientific areas mostly involved in this process are computer science, sociology, psychology and statistics. In the following, we outline the contributions that each of these areas can provide to the design and development of PMIs.

The most effective way to induce a behavioral change in a user ---the goal of moderation interventions--- depends on the personal characteristics and context of the user itself~\cite{drkazkiewicz2022study}. To this regard, \textit{sociology and psychology} can provide the theories and knowledge needed to profile users and design persuasive PMIs. Each user could be described in terms of its social and personality profile~\cite{karanatsiou2022traits}, including its social vulnerabilities~\cite{williams2017individual}. The former characteristics are linked to the emergence of online misbehavior~\cite{kurek2019did}, while the latter are linked to the effectiveness of a given intervention for a given user~\cite{bilewicz2021artificial,hangartner2021empathy}. Knowledge of such individual characteristics is instrumental in the development of effective PMIs.

Then, successful personalization requires accurate user models, powerful analytics and a significant degree of automation to reach scalability. In this regard, many areas of computer science can provide massive contributions to the development of PMIs. For example, the application of theories and practices from \textit{human-computer interaction} (HCI) can drive the process of user modeling. In addition, HCI can contribute to design the way in which moderation interventions are tailored and presented to the users~\cite{cox2016design}, while also leveraging \textit{machine and deep learning} (ML/DL) techniques for obtaining accurate and scalable user representations. For instance, ML/DL can be leveraged to infer personal characteristics of the users based on their publicly available data, such as account information, posting or browsing history, and online social relationships~\cite{mehta2020recent}. Furthermore, ML/DL can also be profitably applied to design multimodal interventions that combine personalized counter-narratives, generated with \textit{natural language processing} (NLP) techniques, and images~\cite{tekirouglu2020generating}. 
In addition, PMIs require to match each user with a favorable intervention. Again, this step can be carried out by resorting to ML/DL techniques and by introducing a new task: estimating the most effective moderation intervention for any given user. The novelty of the task mandates the development of new datasets, sensible baselines, and novel methodologies. At the same time however, its similarity with traditional tasks in the area of \textit{recommender systems} implies that established techniques in that area will likely represent good initial solutions also for the development of PMIs~\cite{lucchese2019learning}.

Finally, validating PMIs and assessing their improvement over generic interventions mandates to go beyond mere correlations and associations, by following rigorous \textit{causal inference} approaches. These could be applied to draw conclusions from survey experiments, or from simulations and field experiments on real online platforms. 
Recently, many of such statistical techniques have been successfully adopted to estimate the effects of generic interventions, such as those techniques designed to detect causal effects in time series data resulting from a given event~\cite{chandrasekharan2020quarantined,horta2021platform,trujillo2022make}. In the future, the same, or similar, techniques could be adapted to evaluate the effects of PMIs.


\section{Challenges and Opportunities}
\label{sec:challenges}
In addition to cleverly combining existing knowledge and to developing new one, PMIs will also mandate solving a number of open challenges. First and foremost, deploying PMIs entails solving a number of \textit{ethical challenges} regarding the use of personal data for user modeling, the right to explanation, and the fairness of automated moderation mechanisms, which must scale in line with sound theory from social and personality psychology. Overcoming such challenges will involve developing agreed-upon ethical standards as well as adopting privacy-preserving computational techniques (e.g., federated learning, anonymization). Motivating the interventions and guaranteeing fair and unbiased decisions will also require the adoption of best practices in explainable and fair recommendation, opening up the opportunity to improve platform transparency and accountability, two areas in which online platforms are being harshly criticized. Some \emph{technical and methodological challenges} are also limiting our capacity to accurately estimate the effects of moderation interventions, such as the difficulty at accounting for confounders and possible exogenous causes~\cite{horta2021platform,trujillo2022make}. In addition, validating new interventions, such as PMIs, involves performing extensive field experiments on online platforms, which typically cannot be carried out without the participation of the platforms themselves, thus more efforts are needed to strengthen collaborations between these and scholars.


Finally, online moderation also carries important \emph{philosophical challenges}. The present proposal embodies the vision to develop the theoretical and technological tools that will enable PMIs. However, the effectiveness of online moderation depends only in part on the availability of powerful and accurate technological tools. Most importantly, it depends on the strategic goals and the regulatory context of the platforms that perform (or not) the moderation. As remarked by Gayo-Avello, the paramount goal of online platforms is to «commoditize and monetize individual communication» and their commitment to scientifically-sound and thorough moderation can only exist to the extent that it does not «affect their investors or the laws under which they operate»~\cite{gayo2015social}. Furthermore, the effects that PMIs will have on the safety and reliability of online platforms will depend on the use that humans will make of them, as it always happens when new technologies are introduced. As such, some actors might use PMIs to manipulate, rather than to persuade, or to censor and silence, rather than to support plurality of opinions and free speech~\cite{gayo2017social}. Overcoming some of these challenges will probably require social, cultural and regulatory changes, in addition to mere technological advancement.

\section{Conclusions}
\label{sec:conclusions}
Personalized moderation interventions (PMIs) promise to transform online moderation by shifting from a coarse-grained, platform-centered approach to a fine-grained, user-centered one. By taking into account the peculiar traits and individual characteristics of the users, PMIs will enable nuanced and effective interventions. Despite this promising outlook, the challenges along this research direction are manifold. Solving them will require combined endeavors from multiple interrelated scientific communities, that we call to join the effort.


\bibliographystyle{ACM-Reference-Format}
\bibliography{references}
\end{document}